\documentstyle[aps]{revtex}
\begin{document}
\draft
\title{Reply to Comment "A posteriori teleportation"}
\author{Dik Bouwmeester, Jian-Wei Pan, Matthew Daniell, \\Harald
Weinfurter, Marek Zukowski, and Anton Zeilinger} \address{Institut
f\"ur Experimentalphysik, Universit\"at, Innsbruck, Technikerstr. 25,
A-6020 Innsbruck, Austria} \maketitle

Braunstein and Kimble observe correctly  that, in the Innsbruck
experiment, one does not always observe a teleported photon
conditioned on a coincidence recording at the Bell-state analyser. In
their opinion, this affects the fidelity of the experiment, but we
believe, in contrast, that it has no significance, and that when a
teleported photon appears, it has all the properties required by the
teleportation protocol. These properties can never be achieved by
"abandoning teleportation altogether and transmitting randomly
selected polarization states" as Braunstein and Kimble suggest. The
fact that there will be events where no teleported photons are created
merely effects the efficiency of the experiment. This suggests that
the measure of fidelity used by Braunstein and Kimble is unsuitable
for our experiment

During the detection of the teleported photons,  no selection was
performed based on the properties of these photons. Therefore, no {\em
a posteriori} measurement in the usual sense as a selective
measurement was performed. The detection of the teleported photon
could have been avoided altogether if we had used a more expensive
detector, p, that could distinguish between one- and two-photon
absorption. The inability of our teleportation experiment to perform
such refined detections does not, however, imply that "a teleported
state can never emerge as a freely propagating state...". Braunstein
and Kimble do not, therefore, reveal a principle flaw in our
teleportation procedure, but merely address a non-trivial practical
question.

\end {document}